\documentclass[11pt,twoside]{article}

\usepackage{asp2006}
\usepackage{epsf}
\usepackage{psfig}
\usepackage{lscape}

\begin{document}

\title{Panel Meeting on Exozodiacal Dust Disks}

\author{O.~Absil,\altaffilmark{1} C.~Eiroa,\altaffilmark{2} J.-C.~Augereau,\altaffilmark{3} C.~A.~Beichman,\altaffilmark{4} W.~C.~Danchi,\altaffilmark{5} D.~Defr\`ere,\altaffilmark{1} C.~V.~M.~Fridlund,\altaffilmark{6} and A.~Roberge\altaffilmark{5}}   

\altaffiltext{1}{Dept.\ AGO, Universit\'e de Li\`ege, 17 All\'ee du Six Ao\^ut, B-4000 Sart Tilman, Belgium}
\altaffiltext{2}{Departamento de F{\'i}sica Te{\'o}rica, C-XI, Universidad Aut{\'o}noma de Madrid, 28049 Madrid, Spain}
\altaffiltext{3}{LAOG--UMR 5571, CNRS and Universit\'e Joseph Fourier, BP 53, F-38041 Grenoble, France}
\altaffiltext{4}{NExScI, California Institute of Technology, 770 S Wilson Ave, Pasadena, CA 91125, USA}
\altaffiltext{5}{Exoplanets and Stellar Astrophysics Laboratory, NASA-GSFC, Greenbelt, MD 20771, USA}
\altaffiltext{6}{ESA-ESTEC, PO Box 299, Keplerlaan 1 NL, 2200AG Noordwijk, The Netherlands}

\begin{abstract}
When observing an extrasolar planetary system, the most luminous component after the star itself is generally the light scattered and/or thermally emitted by a population of micron-sized dust grains. These grains are expected to be continuously replenished by the collisions and evaporation of larger bodies just as in our solar zodiacal cloud. Exozodiacal clouds (``exozodis'') must therefore be seriously taken into account when attempting to directly image faint Earth-like planets (exoEarths, for short). This paper summarizes the oral contributions and discussions that took place during the Satellite Meeting on exozodiacal dust disks, in an attempt to address the following two questions: Do we need to solve the exozodi question? If yes, how to best solve it?
\end{abstract}


\section{Impact of Exozodis on ExoEarth Imaging with Optical Telescopes}

The most obvious effect of exozodiacal dust on optical wavelength direct observations of Earth-like planets is to increase exposure times. Light scattered off the local zodiacal dust and the exozodiacal dust around nearby stars will be mixed with the planet signal in both images and spectra. The larger the final PSF, the more background will be mixed in. This background will likely be the largest source of noise, if the stellar light is canceled by $10^{-10}$ at the planet pixel.

In the background limited case, the exposure time to image a planet is linearly proportional to the exozodi surface brightness. Obviously, if the exozodi is too bright, the integration time to get the required planet SNR becomes prohibitively long. How do various optical mission concepts perform under varying levels of exozodi? Using publicly available information, it is very hard to come up with estimates that can be reasonably compared. In addition, most groups assume a fixed exozodi level for calculating their mission performance, typically one zodi, which is not tremendously informative. There is a strong need for uniform mission performance analysis that pushes to high exozodi levels.

\begin{figure}[!t]
\plotone{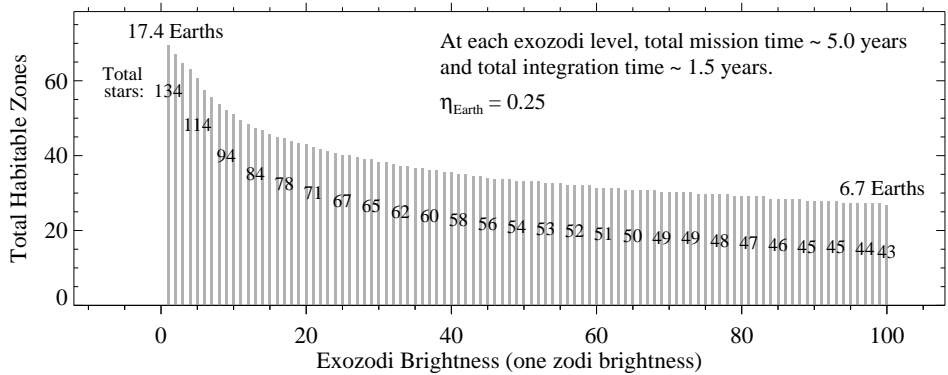}
\caption{Total habitable zones searched vs.\ exozodi surface brightness. Each grey bar represents a possible observing program, assuming $\eta_{\earth} = 0.25$ and the parameters of the \textit{New Worlds Observer} mission concept \citep[for details, see][]{Turnbull10}. The numbers superimposed on some of the grey bars are the total numbers of stars observed in the programs. The y-axis is the total habitable zones searched, which is the sum of the completeness values for all the individual stars observed.  A star's completeness value is the probability that a habitable zone planet would be detected in a single visit, given the possible range of system inclinations and planet eccentricities. The expected number of exoEarths characterized is the total habitable zones searched times $\eta_{\earth}$. To generate a program for each exozodi level, stars were chosen in order of decreasing weighting factor (completeness / exposure time) until the total on-target integration time reached 1.5 years or the total mission time reached 5 years, whichever came sooner. The total integration times were calculated assuming that for $\eta_{\earth} \times 100 = 25 \%$ of the targets, a spectrum with $S/N \geq 10$ and $R = 100$ was obtained in addition to the imaging observation. Total mission times include 11 days per target for moving the occulter.} \label{fig:nwo}
\end{figure}

That being said, Fig.~\ref{fig:nwo} illustrates the decline in mission performance with increasing exozodi brightness for an optical mission. These preliminary calculations were done assuming the parameters of the \textit{New Worlds Observer} mission concept (4-m telescope, broad-band imaging channel covering 500 to 700\,nm), but the general behavior should be similar for other optical wavelength missions using either external occulters or internal coronagraphs. However, it is important to note that these calculations include only statistical errors due to local zodiacal background, exozodiacal background, and unsuppressed starlight. Possible systematic errors associated with modeling light scattered off a non-uniform exozodiacal dust distribution and removing it from a planet image have not been thoroughly characterized for optical missions. Furthermore, confusion between planets and exozodiacal dust structures like resonant clumps remains poorly studied, although initial attempts to evaluate its impact on mission performance appear in \citet{Savransky09} and \citet{Turnbull10}.

\section{Impact of Exozodis on ExoEarth Imaging with IR Interferometry}

The mid-infrared wavelength range presents several advantages for Earth-like planet characterization. In addition to including spectral bands of water, carbon dioxide and ozone, the contrast between a star and an exoEarth is only $\sim 10^7$ whereas it is $\sim 10^{10}$ in the visible. However, resolving the habitable zone around nearby stars in the mid-infrared would require a gigantic telescope with a diameter up to 100\,m. Space-based interferometry is therefore considered as the most promising technique to achieve this goal. A large effort has been carried out during the past decade to define a design that provides excellent scientific performance while minimizing cost and technical risk. This has resulted in a convergence and consensus on a single mission architecture consisting of a non-coplanar X-array, called Emma \citep[see e.g.][]{Cockell09}, using four collector spacecraft and a single beam combiner spacecraft. Such a design enables the implementation of phase chopping, a technique which suppresses from the final output all sources having point-symmetric brightness distributions.

The impact of exozodis on the mission performance is twofold: on one hand, their point-symmetric component contributes to the overall shot noise and can therefore drive the required integration time to detect exoEarths, while on the other hand, asymmetric structures in exozodis (such as resonant clumps) are not suppressed by phase chopping and thereby contribute as possible biases (or false positives), which could prevent the detection of small planets.

\begin{figure}[!t]
\plottwo{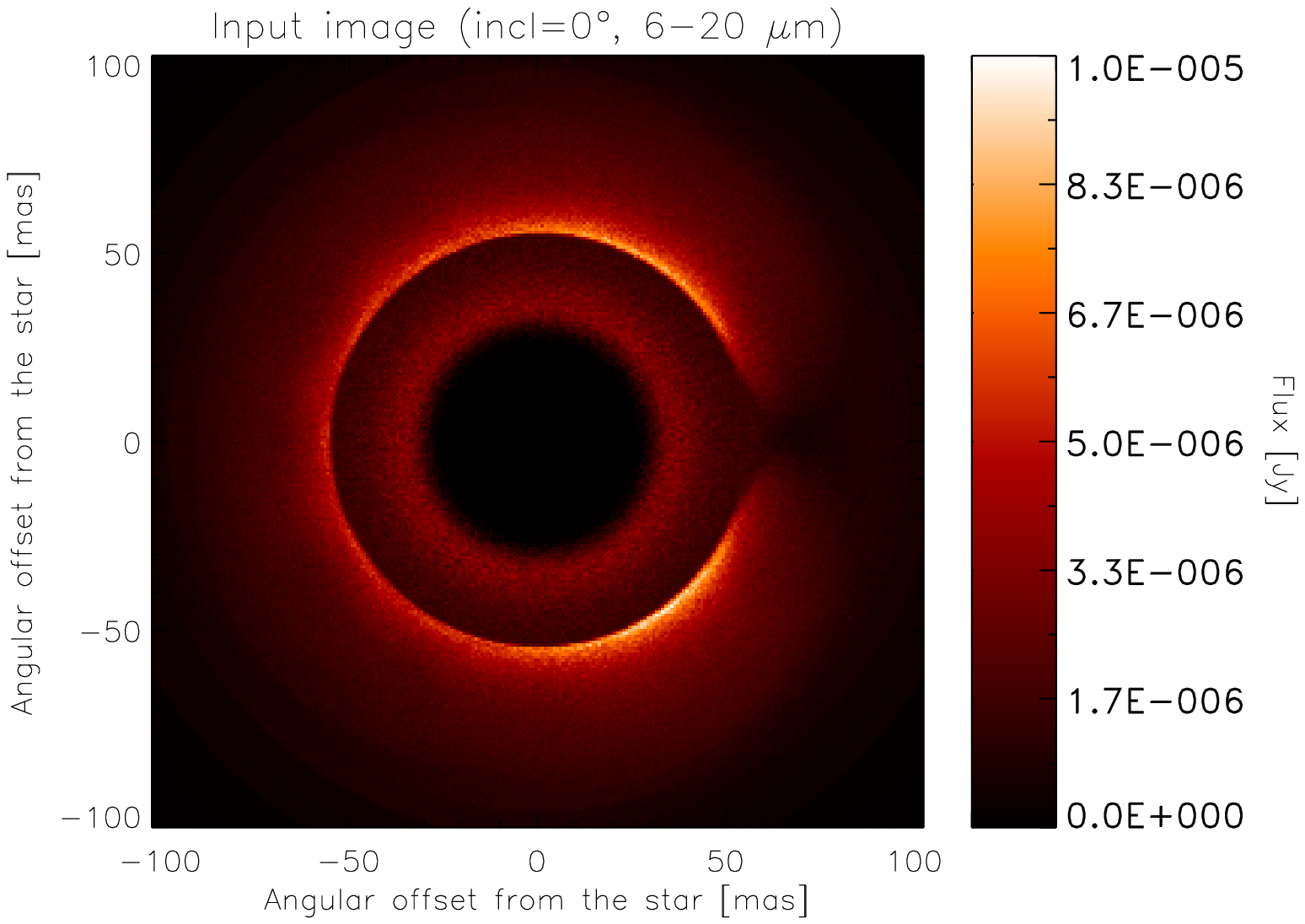}{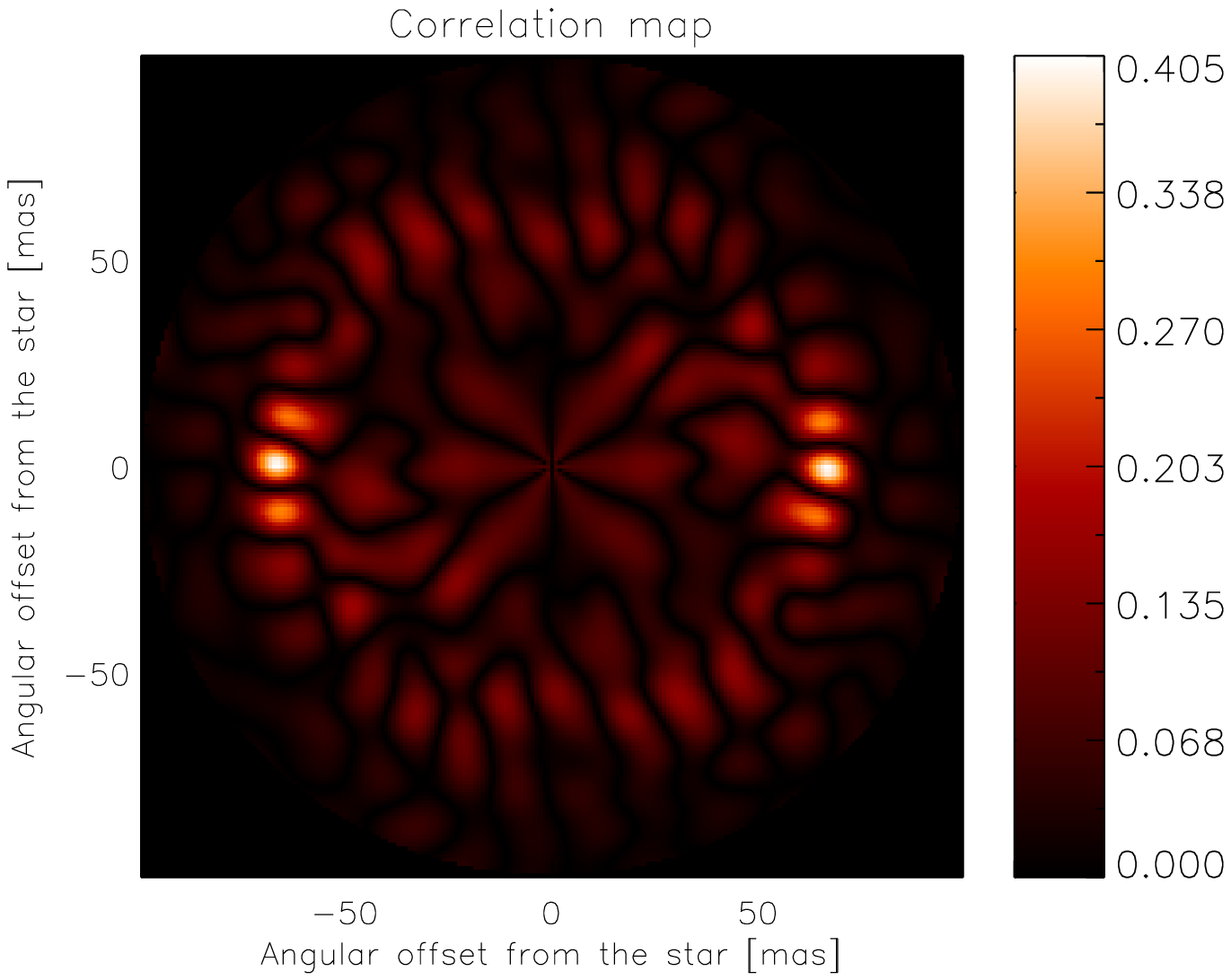}
\caption{\textit{Left:} Simulated image of the thermal flux produced by a 10-zodi face-on exozodiacal dust cloud in the 6-20\,$\mu$m range. The simulation includes the presence of an Earth-mass planet located at 1\,AU on the x-axis, and assumes a Dohnanyi size distribution for the dust grains \citep{Stark08}. \textit{Right:} Corresponding dirty map computed for an Emma X-array nulling interferometer, obtained by cross-correlating the measured signal with templates for the expected signal from a point source at each location on the sky plane \citep{Defrere10}.} \label{fig:disk2map}
\end{figure}

Considering the Emma X-Array design with four 2-m apertures, point-symmetric exozodis of about 100~zodis can be tolerated while preserving 75\% of the mission outcome, i.e., surveying the habitable zone of at least 150~targets with $>90$\% completeness during the 2-yr search program. However, when including the resonant structures created by an exoEarth orbiting at 1\,AU around a Sun-like star (see left part of Fig.~\ref{fig:disk2map}), the asymmetry created by the hole in the dust distribution near the planet significantly contributes to the final detected signal (see dirty map in the right part of Fig.~\ref{fig:disk2map}) and can thereby prevent from detecting the planet itself. The tolerable dust density then goes down to about 15 times the solar zodiacal dust density. This upper limit on the tolerable exozodiacal dust density gives an estimation of the typical sensitivity that precursor instruments will need to reach on exozodiacal disks in order to prepare the scientific program of future exoEarth characterization missions.


\section{Current Exozodiacal Disk Detection Efforts}

Two main techniques are been used to evaluate the amount of exozodiacal dust around main sequence stars. Infrared spectro-photometry can reveal the presence of an excess emission on top of the expected stellar photospheric flux. This requires high-accuracy photometry in the mid-infrared regime, which is generally done from space, e.g., using the various instruments on board the Spitzer Space Telescope. The intrinsic accuracy of Spitzer photometry limits the sensitivity to exozodiacal disks about 1000 times more luminous than the solar zodiacal cloud \citep[assuming the same brightness distribution, see e.g.][]{Bryden06}. Note that, even with an infinite accuracy on the photometry, this technique would still be limited to an accuracy of about $1\%$ by the capability to predict the mid-infrared stellar photospheric flux.

The second way to detect exozodiacal disks overcomes this limitation by angularly separating the signal of the star from its surrounding dust disk. The requested angular resolution can generally be achieved only with infrared interferometry. Two types of interferometers are contributing to exozodi surveys: high-accuracy near-infrared interferometers such as CHARA/FLUOR and VLTI/VINCI \citep[see e.g.][]{Absil06,Absil09}, and mid-infrared nulling interferometers such as MMT/BLINC and the Keck Interferometer Nuller \citep[see e.g.][]{Liu09,Colavita09}. Both types of instruments reach a typical accuracy of $10^{-3}$ on the disk/star contrast, which corresponds to roughly 1000~zodis in the $K$ band and 300~zodis in the $N$ band. These observations are however still restricted to a small amount of targets, due either to limited observing time or to the limiting magnitude of the instrument. A new generation of nulling interferometers, such as the LBTI or the ALADDIN project on the ground, or FKSI in space, could significantly improve the current sensitivity to exozodiacal disks, pushing the detection limit down to $\sim30$~zodi from the ground or $\sim1$~zodi from space (see other paper by Absil et al.\ in this volume).


\section{Current Exozodiacal Disk Modeling Efforts}

\begin{figure}[!ht]
\psfig{figure=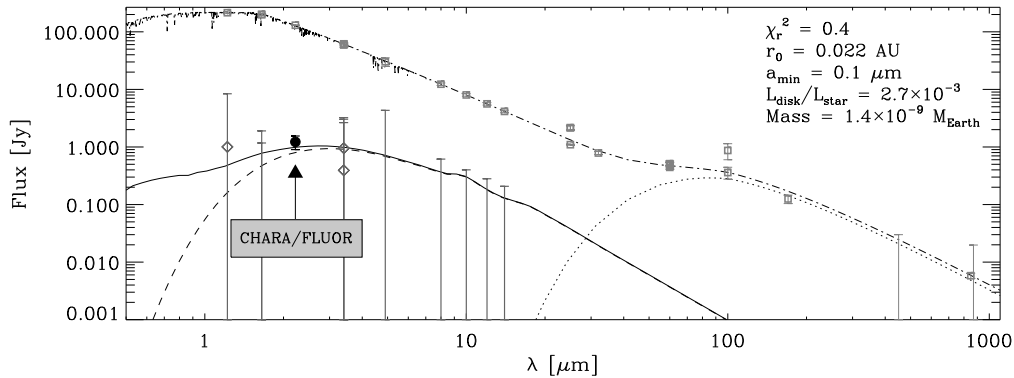,angle=90,height=5cm}
\caption{A possible fit to the photometric excesses (represented with diamonds) and interferometric observations (filled circle at $2.2\,\mu$m) of the $\tau$~Ceti exozodi \citep[from][]{DiFolco07}. Dashed line: thermal emission from the best-fit disk model. Solid line: includes the scattered light contribution. Dotted line: approximative blackbody fit to the long-wavelength excesses produced by a Kuiper-like belt. Dashed-dotted line: total emission from the star-disk system.} \label{fig:taucet}
\end{figure}

The current exozodi modeling efforts follow two main paths. On the one hand, radiative transfer codes for optically thin disks are being employed to reproduce the scarce exozodi measurements, in particular near- and mid-infrared interferometric data (see e.g.\ Fig.~\ref{fig:taucet}), and show that the solar system zodiacal spectrum does not match the spectral energy distributions (SEDs) of detected exozodiacal disks. The fit to the $2.2\,\mu$m excesses with the solar system zodiacal model indeed indicates densities a few thousand times larger than the zodiacal density (e.g., about 3000~zodis for Vega and 5000~zodis for Fomalhaut), but this model predicts much too large flux (by about an order of magnitude) in the mid-infrared compared to the observations. Dust much closer to the star (down to the sublimation radius at a fraction of an AU), and with a sufficient amount of refractory material, are required to shift the spectrum to shorter wavelengths and match the current data sets. Spectral decomposition techniques are also being developed to reproduce extreme Spitzer/IRS spectra showing unusually large amount of warm dust and clear spectral solid-state features \citep[e.g.][]{Lisse09}.

On the other hand, dynamical models employing classical N-body codes are being used to assess the influence of planets on the shape of exozodiacal disks and discuss the dust production mechanisms. Some models for instance simulate the sculpting of an asteroid/dust belt by a planet due to their capture in mean motion resonances \citep[e.g.][]{Stark08}. Larger scale models, involving the outward migration of a planet toward a Kuiper belt are currently being developed, basically relying on the assumption that exozodis are fed by the outer, much more massive disk regions, thereby linking the inner and outer regions of planetary systems. This can either be due to a sudden event, e.g., a Nice LHB-like model \citep{Booth09}, or be more progressive (Augereau et al., in prep.). The amount of dust produced can be assessed for individual systems following this methodology, but it is very sensitive to the assumed planetary system architecture. Finally, collisional models, using statistical approaches, can evaluate the lifetime of asteroid belts due to collisions \citep[e.g.][]{Krivov06,Thebault07}. An ISSI working group has been assembled to further discuss the modeling of exozodis (see \texttt{http://www.issibern.ch/teams/exodust/}).


\section{Conclusions and Recommendations}

This short review illustrates how the final SNR for direct exoEarth detection depends on the quantity of exozodiacal dust around main sequence stars. On one hand, it drives the required integration time to detect the planetary signal as soon as its density reaches a few tens of zodis, and on the other hand, its potential asymmetries induce biases and false positives, which in turn demand the planetary systems to be observed longer in order to extract the actual planetary signal. If space missions had an unlimited lifetime, this wouldn't be a major issue, as one would just skip the inappropriate targets, or integrate longer to eventually reveal their planets. However, space missions are limited in time, and the exozodi issue could thus become a major hurdle in case bright exozodis are common. The sensitivity of current exozodi finders ($\sim 300$~zodis at best) is not appropriate to assess whether exozodis in the 10--100 zodi range are common or not. A dedicated effort to solve this question is therefore mandatory.

Three possible avenues have been identified to make sure that exoEarth imaging missions will be capable of reaching their goals:
\begin{itemize}
\item Perform an exozodi survey with a sensitivity of $\sim 30$~zodis on a statistically meaningful sample of main sequence stars to constrain the distribution of exozodi brightness down to an appropriate level. Space missions can then be designed (in terms of aperture size and mission lifetime) to cope with the inferred mean exozodi level.
\item Measure the exozodi level with an accuracy of $\sim 10$~zodis on all the candidate mission targets, once the targets have been identified (e.g., Sun-like stars hosting exoEarth(s), detected through high-precision astrometric or radial velocity surveys).
\item Use the fact that the statistical distribution of cold debris disks will soon be known at a level similar to the density of the solar Kuiper belt thanks to Herschel. Extrapolating the statistics of cold debris disks towards that of warm exozodis is however not straightforward based on theoretical models, and needs to be backed up by observational data anyway.
\end{itemize}

The Panel therefore recommends that significant efforts and support be put in next generation exozodi finders, starting with LBTI and continuing with mid-term ground-based, balloon-borne or space projects such as ALADDIN or FKSI. The Panel also underscores the importance of continued modeling efforts to better understand the origin and dynamics of second generation dust grains around main sequence stars.


\acknowledgements 

O.A.\ acknowledges the financial support from an F.R.S.-FNRS postdoctoral fellowship, and from the Communaut\'e Fran{\c c}aise de Belgique (ARC -- Acad\'emie universitaire Wallonie-Europe).


\end{document}